\documentclass{comsoc2016}

\usepackage{url, graphicx}

\usepackage{wrapfig}

\usepackage{multirow}
\usepackage[title]{appendix}

\title{A Majoritarian Representative Voting System}
\author{Pietro Speroni di Fenizio and Daniele A. Gewurz}

\begin{document}


\begin{abstract}
We present an alternative voting system that aims at bridging the gap between proportional representative systems and majoritarian, single winner election systems. The system lets people vote for multiple parties, but then assigns each ballot to a single party. This opens a whole range of possible systems, all representative. We show theoretically that this space is convex. Then among the possible parliaments we present an algorithm to produce the most majoritarian result.

We then test the system and compare the results with a pure proportional and a majoritarian voting system showing how the results are comparable with the majoritarian system. Then we simulate the system and show how it tends to produce parties of exponentially decreasing size with always a first, major party.

Finally we describe how the system can be used in a context of a parliament made up of two separate houses.
\end{abstract}


\section{Introduction}

One of the main difficulties in choosing an electoral systems is how to approach the dichotomy between governability and representability. The general consensus is that it is impossible to have a system that is both representative and that assures a sufficient concentration of power in few parties to give them the possibility to efficiently govern.
Such dichotomy will then be reflected in who supports what type of electoral system, with people voting for small parties supporting a representative system (which would give them the possibility of having a presence in parliament), while people voting for bigger parties supporting a majoritarian system (which would give them the possibility to govern). Often the result is a compromise that does not assure neither representability nor governability. These issues are studied in depth in texts such as \cite{reynolds2005} and \cite{sartori1994}.

We are going to show in this paper that this is actually a false dichotomy, and it is possible to generate a system that is a faithful representation of what the electorate desires, and at the same time affords the winning parties enough seats to govern a country.

To do this we shall start by defining under what conditions we shall say that a voting system is representative and, in contrast, what kind of voting systems would fall short of such definition. Then we define our voting procedure and then show that by slightly changing it, we can find many different parliaments all representative of the electorate's wishes. At that point we will have moved our problem from a single solution problem to a problem that admits several solutions. In fact, we will show that the space of representative parliaments is not just a subset of the space of possible parliaments but a convex subset of the simplex of all possible parliaments. Then, it will be possible to design a voting system that, while assuring representability, also satisfies other criteria. Among them, having some larger parties. So, this voting system is a possible definitive solution out of this conundrum.

We tested the system using a simulated scenario made up of 20 parties with 5000 voters. We present the results of this as well. We observed that the resulting sizes of the parties decrease following an exponential law which is independent on the number of voters or on the number of parties. Thus it is always the case that one big party is created followed by all the other parties, decreasing in size always following roughly the same proportion.

We also tested the system by letting two groups of university students vote on the main parties present right now in the Italian parliament. Then we analysed the resulting ballots using this system, as well as using the pure proportional system and the actual Italian system (known as Italicum\footnote{As described in Law nr.\ 52, 6 May 2015, published in \emph{Gazzetta Ufficiale} on 8 May 2015: \url{www.gazzettaufficiale.it/eli/id/2015/05/08/15G00066/sg}}). The result of our system is very similar to the Italicum system, with the added benefit that each elected representative can point to a set of ballots that elected him, something which cannot be done in the Italicum system.


\section{Definition of representativity}

We shall call a system \emph{representative} if:
\begin{enumerate}
\item each voter ballot helped to elect a member of parliament he voted for;
\item each MP has been elected by groups of voters of more or less equivalent size;
\item the only case when a ballot does not help elect a MP is when the group size of the voters is too small for them to elect even a single representative.
\end{enumerate}

In other words, each group of people who is sufficiently big to have at least one representative and sufficiently homogeneous to agree on which party to vote will be represented. At the same time, each elected representative will represent about the same proportion of population.

From this definition the only traditional representative system would be a single constituency proportional system. That is, a system where each party presents a single list, and the parliament is produced by electing people from those lists assigning a number of seats directly proportional to the votes received.

This kind of system is rarely chosen as the resulting parliament is often deemed ``too proportional'', producing a parliament with several small parties and thus unable to govern the country.

This is very well expressed by Florin Fesnic (\cite{fesnic2008}): ``This illustrates the inherent tension that exists in proportional representation systems. If the system is too proportional, it can lead to fragmentation and instability. If the system is made less proportional, either through low district magnitude or through a high threshold, it starts resembling a winner-take-all system, and by doing so defeats the very purpose of proportional representation.''

Let us recall that there are several strategies that are often used to make sure that the resulting system is not too fragmented. 
\begin{description}
\item[Majority Bonus.] This strategy consists in giving to the party with the relative majority extra seats in the parliament, possibly even letting it reach the absolute majority. Of course this system cannot be used inside a representative system, as those seats break the fixed ratio between the elected members and their voters. The extra elected people who gain access to the parliament will not be representative of any specific block of people. And if we were to divide the voters of the first party into smaller blocks this would violate the assumption that each elected candidate should represents blocks of people of the same size.
\item[Election Threshold.] This strategy consists in requiring that the only parties allowed to seat in the parliament are those having at least a certain percentage of votes. In a sense it is inevitable that there is always a threshold, as it is not possible to elect less than one representative. Thus any block of people who chose a party that would elect less than one representative would be automatically excluded from the parliament. The election threshold raises this requirement, thus making the system not representative for all those people who vote smaller parties. Again, this is a strategy that prevents a system from being representative.
\item[Single-Member Constituency Election.] This kind of systems elect a single representative for each local area. This was due historically to the difficulty in organising an election distributed through several geographical areas. So each place would elect its own representative. The problem with this is similar to the problem with the election threshold, but repeated for each constituency. In every constituency only the majority party will be able to send a representative to the parliament. As such any party which is distributed throughout the whole nation, but it is not majoritarian anywhere, will be excluded. This system keeps out of the parliament any except the biggest parties. It is, as such, among the least representative systems. A similar problem, although in a lesser form, occurs for systems with small constituencies. And the bigger a constituency is, the more representative the system will be.
\end{description}

In this paper we will present a system that does not uses any election threshold, uses a single nationwide list, and no majority bonuses, so providing a fully representative system and yet which also provides a big party, able to govern. 

To do this we shall present an alternative way of voting, able to produce in each election multiple results (multiple assignation of the parliamentary seats), each a representative one. Meaning, the voting system will not produce a single result, but a space of possible results, all equally representative. First, in the next two sections, we shall present the most majoritarian-like among those results. And then, in the following two sections, we shall study the general space of all the parliaments that are representative.

\section{Ways to vote}
We shall present here a new voting system. We shall allow each voter to express their preference not for a single party, but for multiple parties. In fact they should vote for all the parties that they feel can represent them, without expressing a preference among those choices. In this the voting ballot will be similar to the one used for (the single winner method) approval voting.

What is important is that expressing multiple options will not be interpreted as splitting the votes between all those parties at the same time, nor for voting for all of them at once. Instead the ballot will still be assigned to one and only one of the parties chosen by the voter. To exactly which party, among those he has chosen, will not be controlled by the voter but by other factors.

It is as if the voter offered a ballot with multiple possibilities while saying: ``any of those parties can represent me''. At this point each ballot will need to be assigned to a single party, and the parliament obtained by associating those ballots to those parties will be appointed. It should be noted that this produces multiple possible parliaments, depending on how the ballots are assigned. But all possible parliaments thus assigned are representative. 

In other words, we can associate to each party a set of ballots (i.e., of people that have voted only for it, or together with other parties), and the number of representatives will be proportional to the number of assigned ballots. So we can associate to each elected representative in the parliament a block of roughly equal size of people that have voted for that representative's party.

Of course, if someone voted for only one party his vote can only be assigned to that party. If each person voted only for one party, the voting system will be equivalent to a purely proportional system. At the other end of the spectrum, if everybody voted for every possible party, then any parliament would be representative.

We shall now see a way to assign the votes to assign to a party the widest possible majority that is representative of people's votes, so making this party more effective in their possibilities to govern. As such, this is the ``most majoritarian'' system among the representative ones.

\section{An application: a majoritarian representative system}
Once people have voted, we need to assign each ballot to one and only one party. Let us suppose we have $p$ parties and $v$ voters.
\begin{itemize}
\item We shall assign to the party that has received the greatest number of votes, party $p_1$, all the ballots from the people that voted for that party, ignoring thus all other preferences they might have chosen. 
\item We are then left with a smaller set of ballots, $v_2$, to assign, consisting of all the ballots in which party $p_1$ does not appear. Again, we will count all the votes in $v_2$, and again choose the party $p_2$ with the most votes to assign to it all the ballots (in $v_2$) voting for it.
\item This leaves a smaller set of ballots $v_3$ to be assigned among the remaining $p-2$ parties.
\item And so on.
\item Once all the ballots have been assigned, the seats are assigned following a purely proportional system.
\end{itemize}

Let us now look at some consequences of this way of assigning the votes. 
\begin{enumerate}
\item First of all, each party will receive a number of ballots in the range between the number of people that voted only for it, and the total number of votes it has received. 
\item Second, when two parties are aiming to represent a similar part of the electorate, the bigger one will absorb all the votes of the people that voted for both. 
\begin{enumerate}
\item So, in the case of two clones\footnote{The notion of ``clones'' in voting theory was introduced by T.N. Tideman in \cite{tideman1987}.}, only one, the bigger, will survive, leaving the other one with no ballots or next to no ballots at all. 
\item On the other hand, two parties that are supported by two disjoint sets of people will not damage each other and will eventually coexist in the parliament.
\end{enumerate}
This should be seen as one of the features of the system, since it identifies ``redundant'' parties and damages or even eliminates them.
\end{enumerate}

Also, we can see this voting method as producing an instability where a small difference among the parties can produce a big change in the result. But this big change is bounded by the resulting parliament having to still be representative. Thus, looking at the results we can always track each party to the set of voters that were willing to be represented by it.

This system can be compared to the Chamberlin-Courant system (see \cite{chamberlin1983}) applied to approval votes (i.e., for the case where each voter specifies who he or she approves), and in particular to the greedy approximation algorithm for the Chamberlin-Courant rule due to Lu and Boutilier in \cite{lu2011} for the case where we have parties of candidates. These procedures become the same as the one presented here if each party list admitted to the parliament is entitled to have a single representative: then the Lu-Boutilier greedy algorithm (or its adaptation to approval voting) starts by selecting the candidate approved most often, elects him, then cancels all voters who voted for him, and then iterates. Instead in our method, at each step, a number of representatives, which is generally larger than 1, is elected.

Indeed, the result obtained electing a committee (with at most one representative per party) or a parliament are very different. Suppose that 52\% of the people agree on a specific candidate/party. If we are electing a committee we would give to this person one seat, and thus a minoritarian position. While if we are electing a parliament (as in our system), we would give to this party 52\% of the seats, so enough to govern alone. Both systems permit representativity, but the Chamberlin-Courant system does not ensure governability.
It should be noted that in a parliament/committee that uses unanimous consensus to govern that difference becomes irrelevant.

\section{Other implementations}\label{other}
The system described in the previous section is just one of the possible applications of the more general framework presented here.

In general, a rule must be explicitly presented to decide the order in which the ballots are assigned to the different parties. This can be an external rule whose actual outcome is decided either after the voting (as in the particular system of the previous section, which assigns seats to parties according the order given by decreasing number of votes received) or beforehand (fixing before the voting the order in which ballots will be assigned).

One way to produce a representative parliament from a set of ballots is to start with a list of parties and then to assign the seats following the order. First assign all the seats that can be assigned to the first party in the list, then among the remaining ballots assigning the seats to the second party in the list, and so on. If we have $n$ possible parties, we would have $n!$ possible combinations. As we shall see in next section, if we consider the convex space of all possible representative parliaments, those $n!$ solutions represent the vertices of this space.

The interior of this space will be represented by parliaments where equal ballots, i.e.\ ballots of people that voted for the same parties, are not all assigned to the same one party. And generally these will be more balanced parliaments, with the size of the parties more similar, but also where no party dominates the parliament.

Another particular case is that in which an external authority fixes this order. For instance, it might be a president or monarch wishing to form a cabinet supported by certain parties.

So this authority begins by assigning ballots to some parties to produce the widest majority \emph{within the limits given by a representative democracy}. As such the system would produce a form of check and balance between the electorate wishes on the one side, and another state power.

Such generalised scenarios are those in which the difference between the system given here and other existing systems is particularly evident, especially the fact that here the ballots express an OR-choice, not an AND-choice. In an approval-voting spirit, voters selecting more than one party are delegating any of them indifferently.
So, if a significant number of voters chooses more than one party, majority coalitions compatible with a given set of ballots and the representativity requirements may differ dramatically and anyone able to choose the order in which the parties should be assigned the ballots would have an enormous power over the produced parliament, but always within the limits that the result would always have to be representative of the voters ballots. And the choice over which party is able to govern with an absolute majority would generally be limited to one or two similar parties.

In principle, we might consider each ballot individually and decide to which list it should be assigned. In practice, it would be unfeasible, but as we shall see in the next section, given a ballot outcome, we may identify a ``space'' of possible parliaments that are compatible with that outcome, so that any point of this space is a representative parliament.

\section{A theoretical result: the space of representative parliaments is convex}
Having considered some special cases of the system we are describing (assigning as many ballots as possible to the most voted party, having an external authority in charge of assigning them etc.), we intend now to study the structure of all the possible solutions (i.e., possible parliaments) for a given an electoral result. In other words, which parliaments are actually representative?

Again, let's assume we have $p$ parties and $v$ voters. We here consider that each voter voted for at least one party and no voter cancelled their ballot. We can always safely bring ourselves into that situation. The space of all the possible parliaments is given by the space of all the $p$-tuples $s = \langle s_1, s_2, \dots, s_p \rangle$ that describe how many votes are assigned to each party; so, $0 \le s_i \le v$. The sum of the $s_i$ must be $v$ (because each vote is counted exactly 1 time). As such, the total space of the possible parliamentary configurations forms a standard simplex, the convex hull of the tuples corresponding to parliaments where a single party has all the seats; geometrically, this is a particular polytope (a generalisation of the notion of 2-dimensional polygons and 3-dimensional polyhedrons).

Let us call this simplex, the space of all the possible parliaments (before divisions and remainders), $S$.

Not all those configurations are representative.
So $R$ will be the subspace of $S$ given by all the configurations that are representative for a given voting result.

Let us now see all the possible ways people can vote. Each person can choose a set of parties, except for the empty set. So each person has $2^p-1$ possible choices. Each of the possible poll results can then be represented by a $v$-tuple $u = \langle u_1, u_2, \dots , u_v \rangle$ with $u_i$ the way voter $i$ voted, i.e.\ one of $2^p-1$ possible choices.

Now, given a particular voting result $u$, we are interested in finding out the shape of $R$.

For each voter we need to choose one of the parties chosen by that voter, that is, we have to choose to which party assign that voter's ballot. This means that if we have $v$ participants, and participant $i$ voted with set $u_i$, we are interested in the Cartesian product of all the sets $u_i$, for all $i$. Let us call this $C$.
 
An example will clarify this. Suppose we have two voters $\{1, 2\}$ and three parties $\{a, b, c\}$. Voter 1 voted for $a$ and $b$ ($u_1 = \{a, b\}$), while voter 2 voted for $b$ and $c$ ($u_2 = \{b, c\}$). We are interested in $C = u_1 \times u_2 = \{a, b\} \times \{b, c\} = \{(a,b), (a,c), (b,b), (b,c)\}$ and each of those elements will represents an acceptable result.

The gist of this section is that the subspace $R$ is convex, that is, for any two points $a$, $b$ in $R$ (i.e., for any to ballot assignation that is compatible with a given voting result) any other point on the line segment having $a$ and $b$ as endpoints is in $R$ (i.e., any parliament in which every party has a number of seats between those it had in $a$ and those in $b$ is itself compatible).

Notice that, even though the points we are interested in have integer coordinates (the numbers of seats assigned to the different parties), the theoretical framework of simplexes and polytopes is relevant, since we may consider the intersection between these geometrical structures and the set of points with integer coordinates.

Consider now, for a given subset $A$ of parties, the set of all voters who chose exactly the set $A$ and denote the number of those voters by $n$. All the $p$-tuples describing the possible assignations of their ballots form, again, a simplex (of dimension $|A|-1$ and edge length $n$). Its extremal points, or vertices, are the $p$-tuples with one coordinate equal to $n$ while the other ones are all zero, corresponding to assigning all $n$ votes to the same party.

To pass from these simplexes, one for each subset of parties, to the more complex structures encompassing all possible ballots, we have to consider all possible sums of elements of the simplexes, one from each simplex. This is called the Minkowski sum of those polytopes, and it is known that the Minkowski sum of polytopes is also a polytope, and precisely the convex
hull of the Minkowski sum of the extremal points of the summands, which were themselves convex, by being simplexes. (For a good survey on definitions and results about polytopes and operations on them, see \cite{weibel2007}.)

Hence, the set of all parliaments obtainable from a given voting result is convex, and its vertices, or extremal points, correspond to the parliaments obtained ``by lists'', that is, as described in section \ref{other}, by fixing in all possible ways an ordering of the parties, and then proceeding to assign the ballots to the parties in that order. Indeed, these are exactly the ways to go ``as far as possible'' in each of the directions represented by the parties.

So we can start, for instance, by taking three possible lists (orderings of the parties) and obtain three possible parliaments: a triangle on the simplex. For each possible parliament inside the triangle, we are sure there exists a possible way to assign the ballots that leads to this parliament.

\section{Testing the system}

We tested the system both through a simulation of an election and by asking two classes of university students, in total 260 students, to vote. We analysed the results by comparing what would have happened if we counted the votes according to the purely proportional system, the latest Italian system, and our majoritarian representative system.

\subsection{Simulating an election}
We tested the system by simulating multiple elections. First we fixed a number of parties and assigned to each party a random position in a 2-dimensional square $[0,1] \times [0,1]$. This is a purely abstract representation of a possible ``space of ideologies'', and as such its dimension might well be different from two (but recall that some authors have actually given descriptions of concrete political positions in term of two axes, as described for instance in \cite{lester1996}).

We then randomly placed, on the same space, 5000 voters. At this point we could define the distance $d$ between each voter and each party as the Euclidean distance between the voter position and the party position. And we thus defined the probability for a voter to vote for a party as $p=(1-d)^k$, for $k$ a fixed parameter.

We then randomly determined each election result and extracted the number of elected representatives for each party. We then ran the election 100 times, randomly placing the parties and the voters each time. And then plotted the average size of the parties, ordered by the biggest to the smallest. The result can be seen in Figure 1.

\begin{figure}[!ht]
  \centering
    \includegraphics[width=\textwidth]{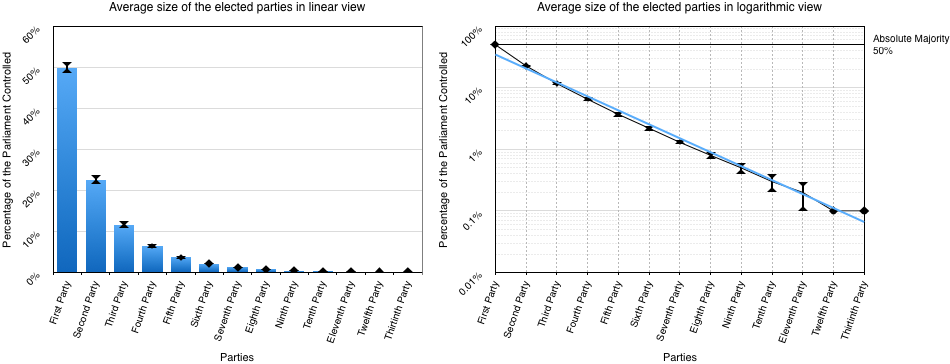}
  \caption{Running the simulation 100 times, with 5000 voters voting 20 parties, and a $k$ of 1.424}
\end{figure}

We tried this for several values of $k$, and several numbers of parties and of voters. Of course we did not know what value of $k$ was a better fit for reality, and future work will try to establish this.

What came out was an extremely consistent result. The party sizes, ordered by number of elected representatives, tended to decrease exponentially, with an exponent that did not depend on the number of parties, nor the number of voters, but did depend on the $k$. For $k = \sqrt{2}$, the first party had an average size of $49.9\% \pm 1.4\%$. That particular value of $k$ was chosen, pending further work, since it seems to approximate the empirical results described in next section.

\subsection{Case report: asking 260 students to vote using this system}
We asked two classes of undergraduates to anonymously vote for the nine main parties in the present Italian political arena. The parties we presented were Partito Democratico (PD), Movimento 5 Stelle (M5S), Forza Italia (FI), Sinistra Ecologia e Libert\`a (SEL), Nuovo Centro Destra (NCD), Fratelli d'Italia, Scelta Civica, Unione di Centro (UDC), Lega Nord (Lega). Each student had to complete 4 different forms:
\begin{description}
\item form 1: students voted for all the parties they would send to the parliament to represent them;
\item form 2: students voted for a single party they would send to the parliament to represent them;
\item form 3: students ordered the parties from the one they liked most to the one they liked least;
\item form 4: students evaluated each party on a 5-option system (terrible, scarce, sufficient, good, excellent).
\end{description}

The first class of students was from the faculty of sociology and consisted of 145 students. The second class was from the faculty of economy and consisted of 115 students. The first thing that should be noted was how inconsistent the votes were. Several students did not vote (in the first form) for the parties they like most (as listed in the third form).

We analysed the results for three groups of people, only the sociologists (group 1), only the economists (group 2), both the sociologists and the economists (group 3).

We then studied the parliament that would be generated by using a purely proportional system, the Italian present election system (the so-called Italicum), and the majoritarian representative election system presented here. The Italicum is a winner-take-all majoritarian system with a threshold of 3\%, and such that if the relative majority party has more than 40\% of the votes it will gain 54\% of the seats, while if the relative majority party has less than 40\% a second election will be held among the two top parties, with the winning party getting 54\% of the seats. The remaining seats are awarded using a proportional system. The only difference between our system and the Italicum is that we used a single constituency, while the Italicum uses 100 constituencies each electing 3 to 9 members of parliament.

So let's see first the winning parliament for the three groups of people -- the sociologists, the economists and all together -- when we use a purely proportional system.
\begin{figure}[!ht]
  \centering
    \includegraphics[width=\textwidth]{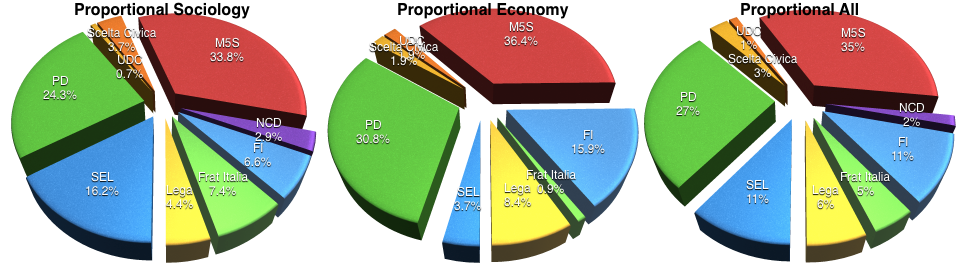}
\end{figure}

And as it could be expected, the result has several parties, with three major parties, but no party able to govern alone. A similar result happened (albeit not using a purely proportional representation) during the 2013 elections, with three very different parties freezing the political situation. Now let us look at the situation using the Italicum.

In no group there was a party that was able to collect 40\% of the electorate. So we had to simulate a runoff voting against the two top parties, PD and M5S. For this we used the third form, where the students could indicate in what order they favoured the parties. In all cases the Movimento 5 Stelle won. So the result showed a clear winner in the Movimento 5 Stelle, to which 54\% of the parliament was given. The rest of the parliament was a bit different between the various groups of people, possibly reflecting the ideological differences between the average person studying economy or studying sociology.
\begin{figure}[!ht]
  \centering
    \includegraphics[width=\textwidth]{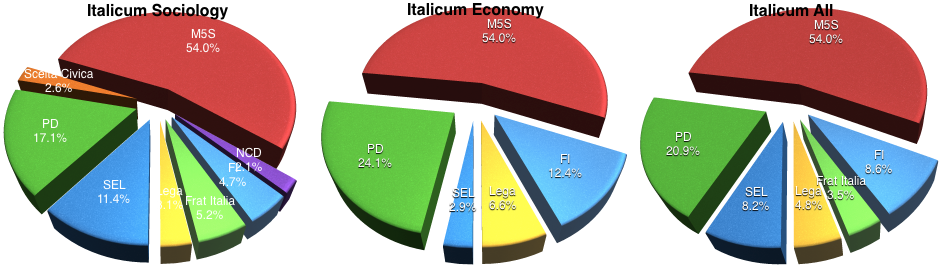}
\end{figure}

Finally we applied our majoritarian representative system to the votes. And the surprising result was, it ended up being very similar to the Italicum.
\begin{figure}[!ht]
  \centering
    \includegraphics[width=\textwidth]{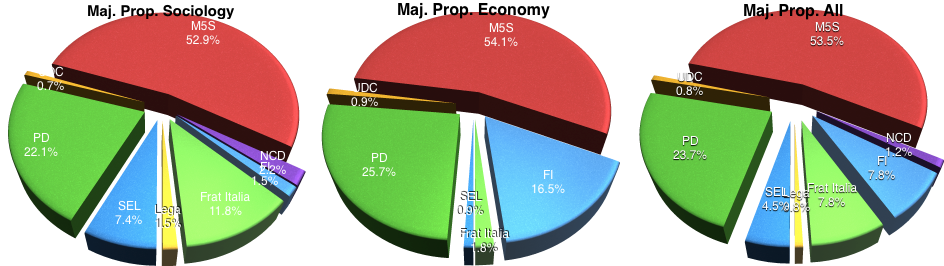}
\end{figure}

In all cases the Movimento 5 Stelle won. Also the percentage was very similar to the one assigned through the Italicum (52.9\%, 54.1\%, 53.5\% versus 54\%).

While there is no reason why the results should be similar to the Italicum, the simulation showed that the size of the first party tended to be always the same, and the two experiments we carried out showed similar results. So if having a first party of 54\% is the optimal percentage to govern, it is a lucky coincidence that this system offers a similar result. But, if this reproduces the behaviour in the simulation, it might be a consistent result and other experiments done with other parties and other voters might also produce a first party with a popularity of around 53\%.

Of course the difference is that it is theoretically possible in the majoritarian representative system to connect each voter to the candidate they helped elect, like in the proportional representation and unlike in the Italicum. Moreover, it is interesting which parties disappeared in the majoritarian representative system. By the way the system is designed, each party will always receive a number of ballots between a minimum of just the ballots of voters that only voted for that party, and a maximum of all the ballots of the voters that also voted for that party. So, when a party disappears, it means that all the voters that voted for it were also happy with other parties. This was the case with Scelta Civica, in all three groups, and Lega Nord and NCD among the economists.

The case of Lega Nord was especially revealing, since it had 19 votes out of 109 valid ballots: so 17.4\% of the voters were willing to also be represented by it, but all of them were also happy to see another party represent them. After Movimento 5 Stelle got his votes the number decreased to 14, after the PD it went to 10, and after Forza Italia to just 1, which was lost when Fratelli d'Italia received its ballots. This example shows very well how the system works: parties that are similar take away votes to each other when one is selected, while parties that are different are barely changed. This is very important in the case of cloning. If a party splits into two clones, only one will generally survive, and it will end up with the original size of the original party. The second, weaker party will only remain (if at all) to represent the people that would not feel represented by the first party.

The complete table of how the votes were assigned in the sociology class clarifies these points even more. Ordering the parties by the number of votes received, we get: M5S, PD, SEL, Fratelli d'Italia, Forza Italia, Lega Nord, Scelta Civica, NCD, UDC. It did not change much after the M5S received its votes. However, after the PD received its votes, the order changed radically and SEL moved from 25 ballots (3rd position) to 12 ballots (4th position) while  Fratelli d'Italia moved from 18 ballots to 16 ballots (3rd position). PD and SEL are in fact quite close in terms of electorate, the former being a center-left party and the latter more decidedly left-wing. Instead Fratelli d'Italia, being on the extreme right, had only 2 ballots in common with the PD. In this way the system, even after assigning ballots to the first party, continued assigning a bigger stake to the second and further parties (if the votes received entitle them to it). So we don't get only a large majority party, but also an opposition made up of few, bigger, parties.

\begin{center}
\begin{tabular}{ |r|r|r|r|r|r|r|r|r|}
\hline
PD & M5S & FI & SEL & NCD & Fr.~Italia & Sc.~Civ. & UDC & Lega \\
\hline
58&{\bf 72}&20&37&6&21&9&3&13\\
{\bf 30}& &11&25&6&18&5&2&6\\
 & &8&12&4&{\bf 16}&4&1&4\\
 & &2&{\bf 10}&3& &4&1&2\\
 & &2& &{\bf 3}& &1&1&2\\
 & &{\bf 2}& & & &1&1&{\bf 2}\\
 & & & & & &0&1& \\
\hline
{\bf 30}&{\bf 72}&{\bf 2}&{\bf 10}&{\bf 3}&{\bf 16}&{\bf 0}&{\bf 1}&{\bf 2}\\
\hline
\end{tabular}
\end{center}

Another thing that should be considered is how should ties be resolved. First of all, in real elections ties are very rare. We did face a tie, and in Appendix A we discuss at length how it was evaluated, and the four possible alternative parliaments that could be reached. In general, recalling what we wrote in section \ref{other}, the simplest way to resolve a tie is to let an external authority decide how to proceed.

\section{Adapting the system to external constraints}
So far, we have discussed the election of a single house of representatives, where we tried to maximise the size of few parties, and voters had no way to express a preference over what politician was elected.

The system can easily be modified to let people express a preference for a politician or another. This can be done by adding a space in the ballot next to each party, where the voter can express their preference by adding the name of a politician. Then the preference is only considered if that ballot is assigned to that specific party. In this way, parties that are elected by a wide majority will have the politicians selected by that majority, while parties that are elected by few people that only voted for them (as the other ballots were assigned elsewhere) will have their representative chosen by those few people.

Another problem has to do with how a parliament made up of two houses is elected, especially since the government might need to be supported by a majority on both houses. In this case it is important that the parties are assigned seats in the same order, especially for the first party. To do this, instead of simply choosing the party that has the maximum number of votes, we need to assign to each party a number which will be the minimum among the number of votes it received in both houses of the parliament. And then pick the party that has the maximum among those numbers. So if we have 5 parties, $A$, $B$, $C$, $D$, $E$, and the number of votes are $A_c$, $A_s$ (the votes for $A$ for the House of Commons and for the Senate, respectively), $B_c$, $B_s$ and so on, then we need to assign $A_m=\min(A_c, A_s)$, $B_m=\min(B_c,B_s)$ etc. And then pick the party that has the highest number among $A_m$, $B_m$, $C_m$, $D_m$, and $E_m$. This will be sure to elect the party with the biggest possible representation in both houses. Once the first party is chosen, the procedure is repeated for the second party using the remaining ballots, and so on.

\section{Conclusions}
We presented a novel electoral system. This system combines two requirements that were previously considered mutually exclusive: producing a parliament that is proportional to what the electoral wants and producing a parliament with one big party able to govern the nation.

\vspace{4mm}
\begin{tabular}{cc|p{4cm}|p{4cm}|}
\cline{3-4}
&& \multicolumn{2}{ c| }{Does it produce one big party able to govern?} \\
\cline{3-4}
&& no & yes \\
\cline{1-4}
\multicolumn{1}{ |c  }{\multirow{2}{*}{Is it proportional?} } &
\multicolumn{1}{ |c| }{no} &  & Italicum and majoritarian systems \\
\cline{2-4}
\multicolumn{1}{ |c  }{}                        &
\multicolumn{1}{ |c| }{yes} & Proportional systems & Majoritarian representative system\\
\cline{1-4}
\end{tabular}
\vspace{4mm}

The system does not always produces a winning party that reaches the absolute majority of the parliament, but just the biggest possible parliamentary presence that would still be proportionally representative. We tested the system on 260 undergraduate students and on a computer simulation. The results with the students appeared to be extremely similar to the result of the Italicum (the Italian majority-bonus system), while retaining the proportional representative quality. On the other hand, the simulation allowed us to gather more data on the relative size of the parliaments that appear, as well as to find out that the sizes of the elected parties tend to decrease exponentially with an exponent that does not depend on the number of parties or the size of the electorate.

All this is possible by letting voters chose multiple parties, and then letting the algorithm choose among those parties which one to assign the ballot to. While we presented a specific algorithm to assign the ballots, many more are possible, and as long as each ballot is assigned to one of the parties voted, all produce a representative system. We proved that the space of all possible representative systems is in fact convex and (if $n$ is the number of parties) its limits are given by the $n!$ ``extreme'' possible ways to assign the ballots. 

One might be struck by the present system as being an ``unfair'' one, since it may lead almost to the disappearance of parties that have a large following, that is, voted by many people (albeit together with other parties). But the point is that this system is ``fair'' towards voters, a larger number of which gets to be represented in any final assignation of the ballot, rather that towards the parties, which are just a tool to categorise the voters' preferences.
 
Another fear that this system might rise is that some party might be so popular that next to everybody votes for it, the result being a party with 90\% of the parliament. This is actually a very unrealistic scenario. Still, it can easily be countered by placing an external limits, and decide that no party will receive ballots that will push it over a threshold (say 55\%) unless the voters voted only for that party. Then we would start to assign the votes starting with the ballots that only voted for that party, and stop once we reach the threshold. A randomly chosen subset of the remaining votes would be assigned to the other parties.

Of course, there remain many questions still open. In particular, it is not clear how voters will react to the idea that they can vote multiple parties but only one will be used and they have no control over which one. Surely some people will react by voting just a single party, but others will welcome the possibility to support the strongest party among the ones they agree with. Also, each voting system does not exist in a vacuum; parties react to it by slowly changing their ideology: moving more toward the centre, or more toward an extreme, depending on what is more favourable. Splitting up, merging or building alliances. All this creates a dynamic feedback loop between the voters and the parties whose results are very difficult to predict. Future works will aim to address some of those questions by trying to develop further the simulation, and measure what values of the parameter $k$ best approaches the willingness of voters to vote for certain parties.

\section{Acknowledgments}
We thank J\'er\^ome Lang for the intellectual support. We also thank Mara Maretti, Lara Fontanella and Luigi Ippoliti, for the interesting discussions and the support in gathering the data, and the three anonymous reviewers for their useful remarks and pointers to literature.

\begin{appendices}
\section{Handling a tie}
A tie can have very different results depending on how it is solved. Such a situation happened when we considered cumulatively all the ballots obtained from the students. The first party to win seats was the Movimento 5 Stelle, then the PD, but then there was a tie. Both Forza Italia and Fratelli d'Italia had 26 ballots left, while the total sum of ballots available between Forza Italia and Fratelli d'Italia was not 52 but 38, since 14 people voted for both Forza Italia and Fratelli d'Italia.

\begin{center}
\begin{tabular}{ |r|r|r|r|r|r|r|r|r|}
\hline
PD & M5S & FI & SEL & NCD & Fr.~Italia & Sc.~Civ. & UDC & Lega \\
\hline
107&{\bf 131}&55&60&13&36&15&13&32\\
{\bf 58}& &31&35&10&30&8&6&20\\
 & &{\bf 26}&13&6&{\bf 26}&4&3&14\\
\hline
\end{tabular}
\end{center}

How should this situation be solved? There are three natural ways in which this situation can be solved, and of course more creative ways are always possible. Each of those three ways has some limits. So rather then offering a solution we shall present all of them. In any case it is obvious that this kind of problems is extremely rare in real elections.

The first possibility is to split the ballots equally among the two parties. This works perfectly if the two parties have no ballots in common. But it somehow defies the purpose of the voting system if there are several ballots in common. After all, if such a tie happened between two opposing big parties, none would win enough seats to govern. In any case the number of ballots to assign and split would be the total number of ballots that voted for both parties, not the sum of the number of ballots that voted for one or the other party. In our case the order in which parties would be assigned seats would be:
\begin{center}
	M5S $>$ PD $>$ FI = Fratelli d'Italia $>$ SEL $>$ NCD $>$ UDC = Lega
\end{center}

The second option is that an external authority -- say, the president of the nation -- would decide which of the two parties should win in this case. And then we continue assigning vote as if that chosen party had naturally won. Note that this might mean that the other party might not be the next one to win the seats. For example in the case above, if after the M5S and the PD we assigned the seats to Fratelli d'Italia then Forza Italia would be the next party. So the order was:
\begin{center}
	M5S $>$ PD $>$ Fratelli d'Italia $>$ FI $>$ SEL $>$ NCD $>$ Lega = UDC
\end{center}

\begin{wrapfigure}{r}{0.6\textwidth}
  \begin{center}
    \includegraphics[width=0.58\textwidth]{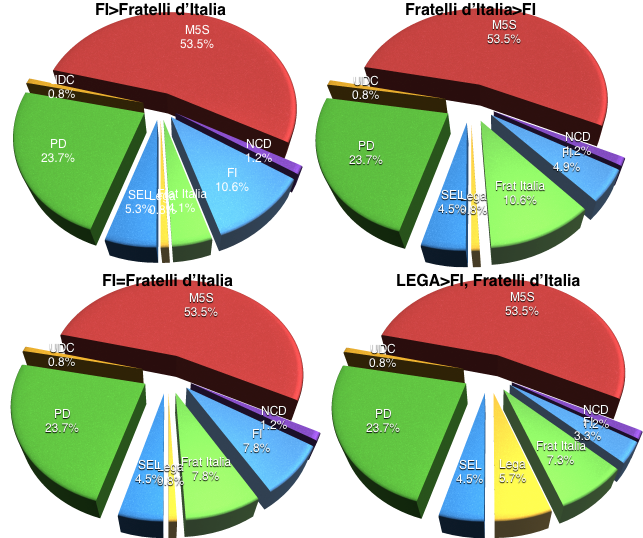}
  \end{center}
  \vspace{-15pt}
  \vspace{-5pt}
\end{wrapfigure}
But if Forza Italia were given the precedence, then the order would have been
\begin{center}
	M5S $>$ PD $>$ FI $>$ SEL $>$ Fratelli d'Italia $>$ NCD $>$ Lega = UDC
\end{center}
since there were more voters in common between SEL and Fratelli d'Italia than between SEL and Forza Italia (which is interesting in itself, considering that SEL represents the extreme left, Fratelli d'Italia represents the extreme right, and Forza Italia represents a more moderate extreme right). 

The final option is to skip temporarily either party and assign the seats to the next party in line, and then go back to the previous parties once the next party has broken the tie. In this case the order in which the parties are assigned would have been:
\begin{center}
	M5S $>$ PD $>$ Lega $>$ Fratelli d'Italia $>$ SEL $>$ FI $>$ NCD $>$ UDC
\end{center}

All this leads to four, substantially different parliaments, all of them representative. And all of them with one party able to form a cabinet, the Movimento 5 Stelle. So all four parliaments are potentially acceptable, but with enormous local differences, with the Lega moving between 5.7\% and 0.8\% and Forza Italia varying between 10.6\% and 3.3\%. Although in this case any of those solution is acceptable, leading to a majoritarian representative parliament, this is not always the case for all possible ties.

In fact, if the tie were for the first place, it is quite likely that any solution that did not privilege either of the two winning parties would lead to a potentially ungovernable situation. So, considering how rare such situations are, a reasonable solution would be to assign to an external authority (like the incumbent president) to act as a tie breaker, choosing what direction should the procedure follow; alternatively, it would be possible to call for a second election just among the first two parties. Of, course, whatever strategy is chosen (showing the tie, breaking the ties using one of these methods, letting a third party benefit from it), it should be defined from before the voting started.

\end{appendices}





\begin{contact}
Pietro Speroni di Fenizio\\
external researcher at\\
Dublin City University\\
Dublin, Ireland\\
\email{2016@pietrosperoni.it}
\end{contact}

\begin{contact}
Daniele A. Gewurz\\
\email{gewurz@gmail.com}
\end{contact}


\end{document}